\documentclass[prl,aps,showpacs,groupedaddress,longbibliography,superscriptaddress,twocolumn,toc=flat,biblatex,footinbib]{revtex4-2}
\usepackage{natbib}
\usepackage[table]{xcolor}
\usepackage[colorlinks=false]{hyperref}
\usepackage[utf8]{inputenc}
\usepackage{color}
\usepackage{bbm} 
\usepackage[english]{babel}
\usepackage{multirow}
\usepackage{amsfonts,amsmath,amssymb,stmaryrd}
\usepackage{braket}
\usepackage{graphicx}
\usepackage{subfigure}  

\usepackage{epsfig}
\usepackage{mathrsfs}
\usepackage{verbatim}
\usepackage{centernot}
\usepackage{ulem}
\usepackage{longtable}
\usepackage{nicefrac}
\usepackage[framemethod=tikz]{mdframed}
\usepackage{siunitx}
\usepackage[nolist]{acronym}

\newcommand{\mw}[1]{\textcolor{black}{#1}}
\newcommand{\pmo}[1]{\textcolor{black}{#1}}
\usepackage{soul}

\usepackage{array}

\usepackage{cancel,ifthen}
\newcommand{\cmnt}[2][NoInPuT]{\ifthenelse{\equal{#1}{NoInPuT}}{}{{\color{red}\sout{#1}}} {\color{blue} #2}}

\usepackage{bm}	
\renewcommand{\vec}[1]{{\bm{#1}}}

\LTcapwidth=\textwidth

\begin{document}

\normalem	

\title{Symmetry-Selective Topological Magnon Engineering by Phonon Angular Momentum}
\author{Markus Wei{\ss}enhofer}
\email[]{markus.weissenhofer@physics.uu.se}
 \affiliation{Department of Physics and Astronomy, Uppsala University, P. O. Box 516, S-751 20 Uppsala, Sweden}

\author{Philipp Rieger}
 \affiliation{Department of Physics and Astronomy, Uppsala University, P. O. Box 516, S-751 20 Uppsala, Sweden}

 \author{Chandan K. Singh}
 \affiliation{Department of Physics and Astronomy, Uppsala University, P. O. Box 516, S-751 20 Uppsala, Sweden}
 
\author{M. S. Mrudul}
 \affiliation{Department of Physics and Astronomy, Uppsala University, P. O. Box 516, S-751 20 Uppsala, Sweden}

\author{Sergiy Mankovsky}
\affiliation{Institute  of  Materials  Science, Technical University of
Darmstadt, 64287  Darmstadt, Germany}
\affiliation{Department of Chemistry/Phys.\ Chemistry, LMU Munich,
   Butenandtstrasse 11, D-81377 Munich, Germany}

\author{Peter M. Oppeneer}
 \affiliation{Department of Physics and Astronomy, Uppsala University, P. O. Box 516, S-751 20 Uppsala, Sweden}

\date{\today}

\begin{abstract}
Dynamical control of Berry curvature remains an outstanding challenge in the engineering of topological phases. Here, we demonstrate control of magnon band structures via coherently driven phonons, based on \textit{ab initio} spin-lattice coupling and Floquet theory. We show that this control is symmetry selective: linearly polarized phonons leave the spectrum unchanged, whereas circular and elliptical phonons carrying finite \ac{PAM} induce chiral interactions that open and tune gaps at Dirac points, generating and reversing topological magnon phases. The gap magnitude and Chern numbers are directly governed by the \ac{PAM}, enabling handedness-selective topology control. Applied to monolayer CrI$_3$, and supported by symmetry analysis, our results establish driven lattice dynamics as a general route to engineering topological bosonic excitations and a versatile platform for Floquet control of magnetism.
\end{abstract}
\maketitle

\begin{acronym}
\acro{PAM}{phonon angular momentum}
\acro{SOC}{spin-orbit coupling}
\acro{DMI}{Dzyaloshinskii-Moriya interactions}
\end{acronym}

\textit{Introduction.}
Berry curvature has emerged as a central concept in modern condensed matter physics, underpinning a wide range of topological phenomena~\cite{Xiao2010,Nagaosa2010,Hasan2010,Qi2011}. It acts as an effective magnetic field in momentum space, giving rise to anomalous transport responses and quantized topological invariants such as Chern numbers. In bosonic systems, and particularly for magnons, Berry curvature plays an equally important role: it governs topological magnon bands, chiral edge modes, and thermal Hall transport~\cite{Matsumoto2011,Matsumoto2014,Onose2010,Mook2014,Mook2014b,Owerre2016,McClarty2022}.

From a symmetry perspective, Berry curvature vanishes identically in the presence of both inversion ($\mathcal{P}$) and time-reversal ($\mathcal{T}$) symmetry. Breaking either symmetry allows for a finite Berry curvature; however, only broken time-reversal symmetry permits a nonzero Chern number, since otherwise the Berry curvature remains antisymmetric in momentum and integrates to zero over the Brillouin zone.

Phonons provide a natural route to dynamically control these symmetries. Linearly polarized phonons, characterized by real polarization vectors and atomic motion along a fixed direction, can break inversion symmetry. In contrast, circularly or elliptically polarized phonons -- where atoms undergo rotational motion around their equilibrium positions -- break time-reversal symmetry~\cite{Zhang2015,Zhu2018}. These modes carry a finite \acf{PAM} defined as $\vec{L}=\sum_i \vec{u}_i \times \vec{p}_i$, with $\vec{u}_i$ and $\vec{p}_i$ being atomic displacements and momenta~\cite{Zhang2014}.

Phonons with finite \ac{PAM} have recently attracted considerable attention due to their interplay with magnetism. They enable ultrafast manipulation of magnetic order, including transient induction of magnetism in paramagnets and multiferroics~\cite{Luo2023,Basini2024}, and reversal of magnetic order in ferrimagnets~\cite{Davies2024}. Conversely, ultrafast magnetic dynamics can excite phonons carrying finite \ac{PAM}, highlighting the reciprocal coupling between spin and lattice degrees of freedom~\cite{Dornes2019,Tauchert2022,Mrudul2025}. These developments point to chiral lattice dynamics as a powerful handle for controlling magnetic properties far from equilibrium~\cite{Juraschek2025}.

Motivated by these considerations, we show in this Letter how driven phonon modes can be used to control magnon band structures. We demonstrate that this control is highly symmetry selective: while linearly polarized phonons leave the magnon spectrum in CrI$_3$ unchanged, phonons with finite \ac{PAM} strongly renormalize the band structure. In particular, they can open or close gaps at Dirac points, thereby inducing or reversing topological magnon phases.

 Our approach combines a recently developed \textit{ab initio} framework for spin-lattice coupling~\cite{Hellsvik2019,Mankovsky2022} with Floquet theory~\cite{Shirley1965,Bukov2015,Eckardt2017,Oka2019,Yarmohammadi2026}, enabling a quantitative description of phonon-driven magnetic excitations. A complementary symmetry analysis based on spin-group theory provides a rigorous understanding of the observed effects and identifies the key role of phonon angular momentum in breaking the relevant symmetries. This establishes a general mechanism for engineering magnon topology via lattice dynamics, which is not limited to CrI$_3$ but is expected to apply broadly to magnetic materials with finite spin-lattice coupling.

\textit{Spin-lattice coupling and Floquet theory.} To describe spin excitations, we adopt an atomistic picture based on the Heisenberg Hamiltonian, which in the absence of the relativistic \ac{SOC} is given by~\cite{Eriksson2017}
\begin{align}
    \hat{\mathcal{H}}    
    &=
    \sum_{ij,ss'}
    \mathcal{J}_{isjs'}
    \hat{\vec{S}}_{is}
    \cdot
    \hat{\vec{S}}_{js'}.
    \label{eq:H_Heis}
\end{align}
Here, $\hat{\vec{S}}_{is(js')}$ are spin operators for spins of length $S_{s(s')}$ located at sublattice $s(s')$ at unit cell $i(j)$, and $\mathcal{J}_{isjs'}$ are the exchange interactions. Note that on-site interactions $\mathcal{J}_{isis}$ are zero.

The exchange interactions $\mathcal{J}_{isjs'}$ are highly sensitive to interatomic distances and local bonding geometries. Therefore, they depend implicitly on the atomic displacements $\vec{u}_{ks''}$ of atoms in the surrounding environment, such that $\mathcal{J}_{isjs'}=\mathcal{J}_{isjs'}(\{ \vec{u}_{ks''}\})$. Importantly, modifications of the exchange interactions are not restricted to displacements of the two atoms directly involved in the exchange path, $(i,s)$ and $(j,s')$. In general, displacements of neighboring atoms 
$(k,s'')$ in the vicinity can also alter the exchange coupling by changing bond lengths, bond angles, and hybridization pathways~\cite{Mankovsky2022}.

The dependence of the exchange interactions on the displacements can be made explicit by performing a Taylor expansion around the equilibrium positions $\vec{r}_{ks''}^0$. To first order, this expansion reads~\cite{Hellsvik2019}
\begin{align}
    \hat{\mathcal{H}}    
    &=
    \sum_{ij,ss'}
    \Big[
    J_{isjs'}
    +
    \sum_{k,s''}
    \vec{A}_{isjs'ks''}
    \cdot
    \vec{u}_{ks''}
    \Big]
    \hat{\vec{S}}_{is}
    \cdot
    \hat{\vec{S}}_{js'}
    ,
    \label{eq:H_SLC}
\end{align}
with the equilibrium exchange interactions $J_{isjs'}=\mathcal{J}_{isjs'}(\{ \vec{u}_{ks''}\equiv0\})$ and the spin-lattice coupling \pmo{parameters} $\vec{A}_{isjs'ks''}=\frac{\partial \mathcal{J}_{isjs'}}{\partial \vec{u}_{ks''}}(\{ \vec{u}_{ks''}\equiv0\})$. 
Recently, efficient methods to calculate spin-lattice coupling parameters from first principles have been developed~\cite{Mankovsky2022,Mankovsky2023,Miranda2024}.

Now we consider a time-periodic displacement of all atoms by a single phonon mode,
\begin{align}
    \vec{u}_{ks''}(t)
    &=
    \frac{A}{2}
    \sqrt{
    \frac{\hbar}{2 m_{s''} \omega_{\vec{q}n}}
    }
    \vec{\phi}_{\vec{q}n,s''}
    e^{i\vec{q}\cdot \vec{R}_k}
    e^{-i\omega_{\vec{q}n}t}
    +
    \mathrm{c.c.}
\end{align}
Here, $\vec{q}$ is the phonon wave vector, $n$ is the phonon band index, $A$ is the (unitless) amplitude, $\vec{\phi}_{\vec{q}n,s''}$ is the (complex) polarization vector, $\omega_{\vec{q}n}$ is the frequency, and $\vec{R}_k$ is the position of the $k$-th unit cell. Inserting $\vec{u}_{ks''}(t)$ into Eq.~\eqref{eq:H_SLC}, we get
\begin{align}
    \hat{\mathcal{H}}  (t) 
    &= \!\!
    \sum_{ij,ss'}
    [
    J_{isjs'}
    +
    A
    e^{-i\omega_{\vec{q}n}t}
    m_{isjs'}
    +
    \mathrm{c.c.}
    ]
    \hat{\vec{S}}_{is}
    \cdot
    \hat{\vec{S}}_{js'},
\end{align}
with 
$
m_{isjs'}
   =
   \frac{1}{2}
   \sum_{k,s''}
    \sqrt{
    \frac{\hbar}{2 m_{s''} \omega_{\vec{q}n}}
    }
    \vec{A}_{isjs'ks''}
    \cdot
    \vec{\phi}_{\vec{q}n,s''}
    e^{i\vec{q}\cdot \vec{R}_k}
$
being the coefficients of the time-dependent corrections to the exchange interactions.  They obey the symmetry $m_{isjs'}=m_{js'is}$. \pmo{Note} \mw{that these corrections do not modify the magnetic ground-state energy, since their time average vanishes for any static spin configuration.} This Hamiltonian is time-periodic, $\hat{\mathcal{H}}(t)=\hat{\mathcal{H}}(t+T)$, with period $T=\frac{2\pi}{\omega_{\vec{q}n}}$.


Floquet theory~\cite{Oka2019} provides a convenient way to analyze time-periodic systems. If the phonon frequency is larger than those of magnons in the system, we can use the van Vleck expansion to obtain an effective, time-independent Hamiltonian~\cite{Bukov2015},
\begin{align}
    \begin{split}
    \hat{\mathcal{H}}_\mathrm{eff}
    &=
    \hat{\mathcal{H}}_0
    +
    \sum_{m\neq 0}
    \frac{
    [ \hat{\mathcal{H}}_{-m},\hat{\mathcal{H}}_{m} ]
    }{2m\hbar\omega_{\vec{q}n}}
    +
    \mathcal{O}(\omega_{\vec{q}n}^{-2}),
    \label{eq:Heff_vV}
    \end{split}
\end{align}
with $\hat{\mathcal{H}}_{m}=\frac{1}{T} \int_0^T \mathrm{d}t\  e^{-im\omega_{\vec{q}n}t} \hat{\mathcal{H}}(t)$ being the Fourier components of the Hamiltonian. Here, only Fourier components with $m\in\{-1,0,1\}$ are finite, with $\hat{\mathcal{H}}_0$ being the Heisenberg Hamiltonian without displacements. A short calculation (details in the SM~\cite{SM}) gives the lowest order correction term, 
\begin{align}\hat{\mathcal{H}}_\mathrm{eff}^{(1)}
    =
    4
     i
     |A|^2 \!\!\!
    \sum_{ijk,ss's''} \!\!\!
    \frac{m_{isjs'}(m_{isks''})^*}{\hbar\omega_{\vec{q}n}}
    \hat{\vec{S}}_{is}
    \cdot
    (
    \hat{\vec{S}}_{js'}
    \times
    \hat{\vec{S}}_{ks''}
    ),
\end{align}
which describes a three-spin interaction proportional to the scalar spin chirality $\hat{\vec{S}}_{is}\cdot (\hat{\vec{S}}_{js'}    \times\hat{\vec{S}}_{ks''})$. Such a term has been found to emerge for light-induced magnetic interactions~\cite{Yambe2023}, but so far has not been discussed in the context of phonon-driven manipulation of magnetic interactions. The next term in the expansion $\hat{\mathcal{H}}_\mathrm{eff}^{(2)}$ is proportional to $\omega_{\vec{q}n}^{-2}$ and describes a four-spin interaction (see SM \cite{SM}, for details).

Performing the Holstein-Primakoff transformation~\cite{Holstein1940} of the terms in $\mathcal{H}_\mathrm{eff}$ up to linear order in $\omega_{\vec{q}n}^{-1}$, and keeping terms of quadratic order in the magnon operators $\hat{a}_{\vec{k}s}^{(\dagger)}$, we get \cite{SM}
\begin{align}
    \begin{split}
    \hat{\mathcal{H}}
    \hspace{-0.125em}
    &=
    \hspace{-0.25em}
    \sum_{\vec{k}ss'}
    \hspace{-0.2em}
    \big(
    \tilde{J}_{\vec{k}ss'}
    \hspace{-0.1em}
    +
    \hspace{-0.1em}
    |A|^2
    (
    \tilde{m}_{\vec{k}ss'}
    \hspace{-0.1em}
    -
    \hspace{-0.1em}
    \tilde{m}_{-\vec{k}s's}
    )
    \big)
    \hat{a}_{\vec{k}s}^\dagger
    \hspace{-0.1em}
    \hat{a}_{\vec{k}s'}
    \hspace{-0.1em}
    -
    \hspace{-0.1em}
     \tilde{J}_{\vec{0}ss'}
    \hat{a}_{\vec{k}s}^\dagger
    \hspace{-0.1em}
    \hat{a}_{\vec{k}s}.
    \label{eq:H_mag}
    \end{split}
\end{align}
Note that we have chosen an expansion around a ferromagnetic state with equal magnetic moments at each sublattice (since below we will apply the framework to CrI$_3$) and the summation over $s,s'$ is only over magnetic sublattices. We want to emphasize that magnon Hamiltonians for any other magnetic state (e.g., antiferro-, altermagnets, or non-collinear states) can also be derived from $\hat{\mathcal{H}}_\mathrm{eff}$ in a similar manner.

The coefficients to the magnon Hamiltonian are
\begin{align}
    \tilde{J}_{\vec{k}ss'}
    &=
    2
    S
     \sum_{j}
    J_{is,js'}
    e^{i\vec{k}\cdot(\vec{r}_j-\vec{r}_i)} ,
    \\
    \begin{split}
   \tilde{m}_{\vec{k}ss'}
    &=
    \frac{4 S^2}{\hbar\omega_{\vec{q}n}}
    \hspace{-0.2em}
    \sum_{jk,s''}
    \hspace{-0.2em}
    m_{is''ks}(m_{is''js'})^*
    e^{i\vec{k}\cdot(\vec{r}_j-\vec{r}_i)}
    e^{-i\vec{k}\cdot(\vec{r}_k-\vec{r}_i)}
    \\
    &+
    (
    m_{isjs'}(m_{isks''})^*
    -
    m_{isks''}(m_{isjs'})^*
    )
    e^{i\vec{k}\cdot(\vec{r}_j-\vec{r}_i)} 
    \end{split}
\end{align}
and can be directly computed from the exchange interactions and spin-lattice coupling parameters.

\begin{figure}
    \centering
    \includegraphics[width=1.0\linewidth]{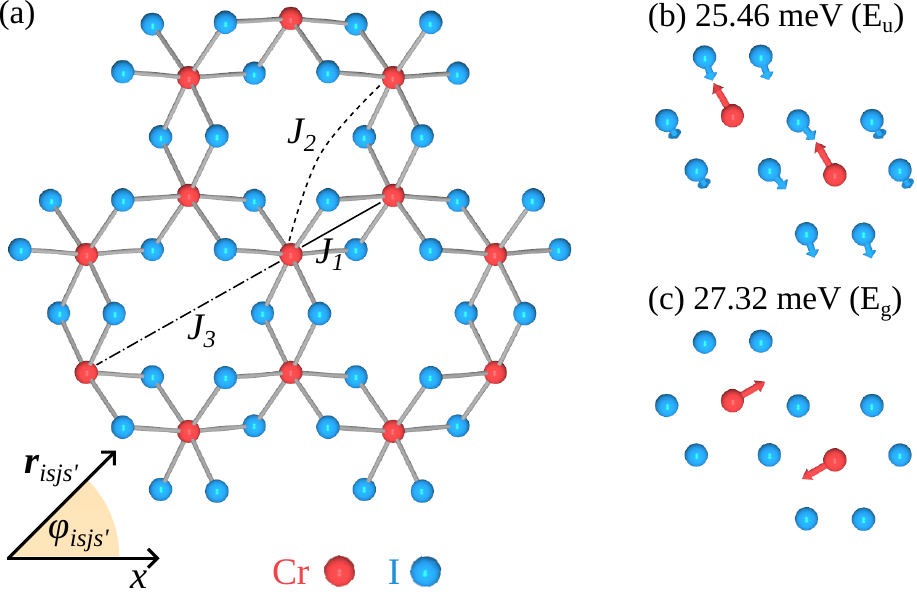}
    \caption{(a) Top view of the atomic lattice of monolayer CrI$_3$. $J_1$ to $J_3$ denote the nearest- to third-nearest-neighbor exchange interactions between magnetic Cr atoms. (b) and (c) Schematic representations of one representative eigenvector from each pair of analyzed degenerate phonon modes with energies and group representations as labeled.}
    \label{fig:CrI3}
\end{figure}

\textit{Results for monolayer CrI$_3$.}
For our study we consider CrI$_3$, which is a layered magnetic van der Waals material exhibiting ferromagnetic order with a Curie temperature of $\SI{61}{\kelvin}$ in bulk crystals and $\SI{45}{\kelvin}$ in the monolayer limit~\cite{Huang2017}. The bulk (monolayer) crystal structure belongs to the symmetry group $R\bar{3}$ ($P\bar{3}1m$), and the magnetic lattice consists of two distinct chromium sublattices forming a honeycomb network, with localized moments characterized by spin $S$.
Inelastic neutron scattering revealed that the energies of magnons in CrI$_3$ go up to approximately $\SI{20}{\milli\electronvolt}$ \cite{Chen2018}. We compute the magnetic exchange interactions $J_{isjs'}$ and the spin-lattice coupling parameters $\vec{A}_{isjs'ks''}$ using the relativistic formulation of the so-called LKAG (Liechtenstein-Katsnelson-Antropov-Gubanov) formalism
\cite{Liechtenstein1987,Ebert2009} and its extension for spin-lattice coupling~\cite{Mankovsky2022}, implemented in the SPRKKR package~\cite{Ebert2011,SM}. Among the dominant interactions, we obtain nearest-neighbor, next-nearest-neighbor, and third-nearest-neighbor exchange couplings of magnitude 
$J_1/S^2=\SI{3.027}{\milli\electronvolt}$, $J_2/S^2=\SI{0.538}{\milli\electronvolt}$ to $J_3/S^2=\SI{-0.048}{\milli\electronvolt}$, respectively (see Fig.~\ref{fig:CrI3}), with $S=3.2/2$. 

The phonon properties are obtained from density functional theory calculations performed using VASP~\cite{Kresse1996a,Kresse1996b} and phonopy~\cite{Togo2023}. We focus on degenerate optical phonon modes at the $\Gamma$ point. To avoid resonant magnon-phonon processes, only phonons with energies well above those of the magnons are considered. In particular, we analyze two pairs of symmetry-protected degenerate modes, illustrated in Fig.~\ref{fig:CrI3}(b) and (c), belonging to the $E_u$ and $E_g$ irreducible representations. \mw{Direct off-resonant excitation of magnons by a THz laser resonant with any of the considered phonon modes is negligible \cite{SM}.}

Circularly polarized phonon modes are constructed from the degenerate linear polarization vectors according to $\vec{\phi}_{\vec{q}n,s}^\pm=(\vec{\phi}_{\vec{q}n,s}^{(1)}\pm i \vec{\phi}_{\vec{q}n,s}^{(2)})/\sqrt{2}$ which correspond to clockwise (+) and counter-clockwise ($-$) rotational lattice vibrations.

First, we calculate $m_{is,js'}$, the time-dependent corrections to the exchange interactions, induced by circularly polarized phonons. In general, we find that their absolute value is the same for all neighbors in a certain shell; they only differ by a phase. We find the phase difference between the induced exchange of two pairs of neighbors to be twice the angle enclosed by the respective bond vectors. That is why we can write $m_{is,js'}=m_{l}e^{\mp i 2\varphi_{isjs'} + i\alpha_{l}}$, with $l=1,2,3,\dots$ indexing the neighbor shells (analogous to the $J_{l}$; see Fig.~\ref{fig:CrI3}), $\varphi_{isjs'}$ being the bond angle and $\alpha_{l}$ some constant phase for each shell. We find ``$-$" in the exponential for clockwise circular phonons and ``$+$" for counter-clockwise circular phonons. The absolute values for the first three neighbor shells are listed in Table~\ref{tab:mij}.

Interestingly, the $E_u$ circular modes only modify the intrasublattice exchange (via $m_2$) and do not affect the intersublattice exchange. For $E_u$ phonons, the displacements are odd under inversion and Cr sublattice exchange. Since we consider only terms up to linear order in displacements in Eq.~\eqref{eq:H_SLC}, the $m_{is,js'}$ are also odd under this transformation. Thus, it follows that $m_{i1,j2}=-m_{j2,i1}$, which, together with $m_{is,js'}=m_{js',is}$, implies that the intersublattice exchange $m_{i1,j2}$ is zero.
From the $m_{isjs'}$, we obtain the coefficients $\tilde{m}_{\vec{k}ss'}$ for the magnon Hamiltonian~\eqref{eq:H_mag}. The calculation of its eigenenergies and eigenmodes is performed numerically using Colpa's method~\cite{SM,Colpa1978}.

\begin{table}[t!]
    \centering
    \caption{Absolute values of time-dependent corrections to the exchange interactions $m_{is,js'}$ in the first three neighbor shells induced by circular phonon modes constructed from two pairs of degenerate phonon modes with energies and symmetries as labeled. All values are in $\si{\milli\electronvolt}$. }
    \begin{tabular}{c|c|c|c|c}
         \hline
         phonon energy & symmetry & $m_1/S^2$ & $m_2/S^2$ & $m_3/S^2$ \\
         \hline
         $\SI{25.46}{}$ & $E_u$ & $0$ & $\SI{0.019}{}$ & $0$\\
         $\SI{27.32}{}$ & $E_g$ & $\SI{0.154}{}$ & $\SI{0.012}{}$ & $\SI{0.011}{}$\\
         \hline
    \end{tabular}
    \label{tab:mij}
\end{table}

\begin{figure}
    \centering
    \includegraphics[width=1.0\linewidth]{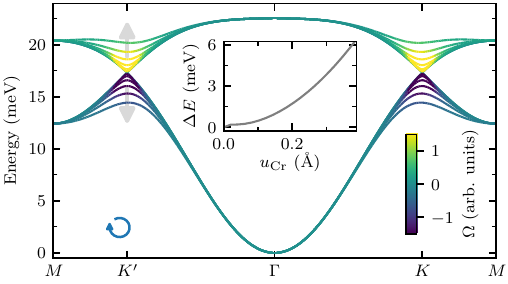}
    \caption{Engineering of magnon bandstructure of CrI$_3$. Arrows indicate the opening of a gap at the Dirac points $K$ and $K'$ upon driving by clockwise circular $E_g$ phonon mode and the color coding shows the Berry curvature of the magnon bands. The inset shows the magnon energy gap versus Cr displacement.}
    \label{fig:dispersion}
\end{figure}

Fig.~\ref{fig:dispersion} illustrates the modification of the magnon spectrum induced by clockwise circular $E_g$ phonon modes. In the absence of phonon driving, the two magnon bands exhibit Dirac-like crossings at the $K$ and $K'$ points. For a finite circular phonon amplitude, these degeneracies are lifted and energy gaps open at both high-symmetry points. The magnitude of the induced gaps scales linearly with $|A|^2$, and thus quadratically with the displacement amplitude of the Cr atoms from their equilibrium positions (see inset of Fig.~\ref{fig:dispersion}).

The gap opening signals a topological phase transition. For any finite drive amplitude, the magnon bands acquire nontrivial Chern numbers $C_n=\frac{1}{2\pi}\int \Omega_{\vec{k}n} \mathrm{d}^2k$ where the Berry curvature is defined as $\Omega_{\vec{k}n}=i (\braket{\partial_{k_x}n_\vec{k}|\partial_{k_y}n_\vec{k}}-\braket{\partial_{k_y}n_\vec{k}|\partial_{k_x}n_\vec{k}})$~\cite{Xiao2010}. We obtain $C_n=\pm 1$ for the two bands~\cite{SM}, demonstrating that the circular-phonon-induced gap is topological in nature. 
Reversing the sense of atomic rotation to counter-clockwise inverts the sign of the Berry curvature throughout the Brillouin zone and consequently flips the Chern numbers.

The dependence of the topological gap opening on the phonon properties, including the \ac{PAM}, can be made explicit by considering a superposition of the two degenerate linear polarization vectors, $\vec{\phi}_{\vec{q}n,s}(\psi)=(\vec{\phi}_{\vec{q}n,s}^{(1)} + e^{i\psi}\vec{\phi}_{\vec{q}n,s}^{(2)})/\sqrt{2}$, where $\psi$ captures the polarization: for $\psi=0,\pm\pi$ the atomic motion is linearly polarized, whereas $\psi=\pi/2$ corresponds to clockwise circular motion and $\psi=-\pi/2$ to counter-clockwise circular motion.

Refs.~\cite{Zhang2014,Weissenhofer2025} showed that the \ac{PAM} of a phonon mode is related to its polarization vectors through $\vec{L}_{\vec{k}n}= i\hbar\sum_{s}\vec{\phi}_{\vec{q}n,s}\times(\vec{\phi}_{\vec{q}n,s})^*$. For the parametrization above, this expression simplifies to    $ \vec{L}_{\vec{k}n}=  \hbar \sin \psi  \sum_{s'} \vec{\phi}_{\vec{k}n,s'}^{(1)}  \times \vec{\phi}_{\vec{k}n,s'}^{(2)}$.

\begin{figure}
    \centering
    \includegraphics[width=1.0\linewidth]{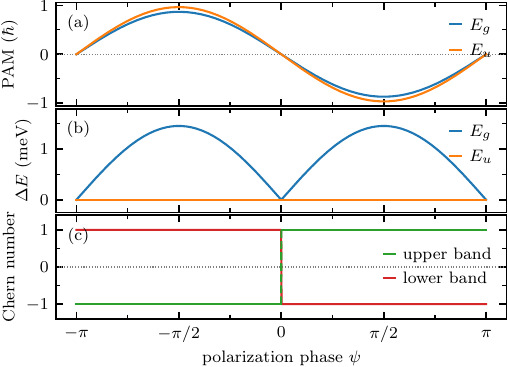}
    \caption{\ac{PAM}-selective control of magnon gap and topologies in the absence of \ac{SOC}. (a) \ac{PAM} and (b) magnon gap at $K,K'$ (for $u_\mathrm{Cr}{\color{black}=|A|
    \sqrt{
    \frac{\hbar}{2 m_\mathrm{Cr}\omega_{\vec{q}n}}
    }
    |\vec{\phi}_{\vec{q}n,\mathrm{Cr}}|}=\SI{0.184}{\angstrom}$) versus phonon polarization for $E_g$ and $E_u$ modes. (c) Chern numbers of both magnon bands versus phonon polarization upon driving by $E_g$ modes (results for $\psi =\pm \pi$ are not shown).}
    \label{fig:gap_vs_ellipticity}
\end{figure}

This sine-like behavior is confirmed in Fig.~\ref{fig:gap_vs_ellipticity}(a), which shows the \ac{PAM} for $\psi\in(-\pi,\pi)$ for both pairs of degenerate phonon modes considered here. The maximum \ac{PAM}, given by $\hbar\sum_{s'}\vec{\phi}_{\vec{k}n,s'}^{(1)}\times\vec{\phi}_{\vec{k}n,s'}^{(2)}$, is slightly smaller for the $E_g$ modes.

Figures~\ref{fig:gap_vs_ellipticity}(b) and (c) further reveal that the magnon gaps $\Delta E$ at $K$ and $K'$ induced by $E_g$ phonons scale as $|\sin\psi|$, while the corresponding Chern numbers follow $\pm\mathrm{sgn}(\psi)$. This establishes the relations
\begin{equation}
\Delta E \propto |\mathrm{PAM}|, \qquad C_n \propto \pm\mathrm{sgn}(\mathrm{PAM}),
\end{equation}
which constitute a central finding of this work: the gap and topology of magnon bands can be controlled by the \ac{PAM} of lattice vibrations.

The absence of any modification of the magnon dispersion under $E_u$ phonon driving originates from the lack of induced intersublattice interactions for these modes (see Tab.~\ref{tab:mij}). Such interactions are required for finite $\tilde{m}_{\vec{k}ss'}-\tilde{m}_{-\vec{k}s's}$, and their absence therefore prevents the opening of a gap in the magnon spectrum (details in the SM \cite{SM}).

To understand why circular or elliptical phonons can open a gap \mw{through nonrelativistic spin-lattice coupling} while linearly polarized phonons cannot, we analyze the symmetry constraints on $\tilde{m}_{\vec{k}ss'}$ imposed by the magnon Hamiltonian~\eqref{eq:H_mag}. We first rewrite the magnon Hamiltonian in Bogoliubov--de Gennes form~\cite{Mook2019}, $\hat{\mathcal{H}}=\frac{1}{2}\sum_{\vec{k}ss'} \hat{\vec{X}}_{\vec{k}s}^\dagger \mathcal{H}_{\vec{k}ss'}\hat{\vec{X}}_{\vec{k}s'}  $ where $\hat{\vec{X}}_{\vec{k}s}^\dagger = (\hat{a}_{-\vec{k}s},\hat{a}_{\vec{k}s}^\dagger)$ and $\mathcal{H}_{\vec{k}ss'}=(\mathcal{H}_{\vec{k}ss'}^{0} -|A|^2\Delta \mathcal{H}_{\vec{k}ss'})$.
The $2\times2$ matrices $\mathcal{H}_{\vec{k}ss'}^0 =  (\tilde{J}_{\vec{k}ss'}-\tilde{J}_{\vec{0}ss'}\delta_{ss'}) \sigma_0$ and $\Delta \mathcal{H}_{\vec{k}ss'}= 2\tilde{m}_{\vec{k}ss'} \sigma_3$ describe the equilibrium interactions and the phonon-induced corrections, respectively. Here $\sigma_i$ with $i\in\{0,1,2,3\}$ denote the identity and the three Pauli matrices~\cite{PauliMatrices}.

The Bogoliubov--de Gennes structure implies a generalized particle--hole symmetry requiring $\sigma_1 \mathcal{H}_{\vec{k}ss'}\sigma_1 = \mathcal{H}_{-\vec{k}ss'}^*$.
Consequently, $\Delta\mathcal{H}_{\vec{k}ss'}=-\Delta\mathcal{H}_{-\vec{k}ss'}^*$ independent of which phonon mode is excited. In the absence of phonons carrying \ac{PAM}, and neglecting \ac{SOC} (see below), the system is additionally invariant under combined time reversal and spin reversal, implying $\Delta\mathcal{H}_{\vec{k}ss'}=\Delta\mathcal{H}_{-\vec{k}ss'}^*$~\cite{Weissenhofer2026}. The two symmetry relations therefore enforce $\Delta\mathcal{H}_{\vec{k}ss'}=0$
for linearly polarized modes. Circular or elliptical phonons, however, carry finite \ac{PAM} and thus break this effective time- and spin reversal symmetry for magnons,  allowing for a finite $\tilde{m}_{\vec{k}ss'}$ and the opening of a topological magnon gap.

\begin{figure}
    \centering
    \includegraphics[width=1.0\linewidth]{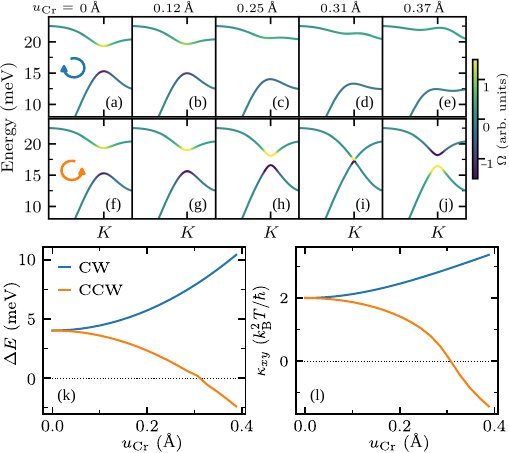}
    \caption{Engineering of magnon bandstructure of CrI$_3$ in the presence of \ac{SOC}, assumed to manifest in \ac{DMI}. \mw{(a--e) and (f--j)} show the magnon energies around $K$ driven by clockwise (CW) and counter-clockwise (CCW) $E_g$ phonons \mw{with Cr displacements $u_\mathrm{Cr}$ as labeled}. \mw{T}he color coding \mw{indicates} the Berry curvature. \mw{(k) and (l)} depict the magnon energy gap and the anomalous thermal Hall conductivity versus Cr displacement.    
    }
    \label{fig:dispersion_DMI}
\end{figure}

According to experiments, the magnon spectrum of CrI$_3$ is already gapped in the absence of phonon driving~\cite{Chen2018}. 
The microscopic origin of this gap remains under debate -- ranging from Kitaev-type interactions \cite{Lee2020,Xu2018} to \ac{DMI}~\cite{Chen2018,Kartsev2020,Chen2021}.  We adopt the latter scenario in the following. Nearest-neighbor \ac{DMI} is forbidden by inversion symmetry, and we therefore include next-nearest-neighbor \ac{DMI} with a strength taken from Ref.~\cite{Chen2018}. As a \ac{SOC} effect, \ac{DMI} breaks the combined time-reversal and spin-reversal symmetry discussed above, which explains the presence of a finite gap already in equilibrium (see Figs.~\ref{fig:dispersion_DMI}(a) and (\mw{f})).

Upon driving the system with circular phonons, the gap can be tuned in a handedness-selective manner: clockwise phonons enhance the gap, whereas counter-clockwise phonons reduce it, eventually closing and reopening it with inverted topology, i.e., with exchanged Chern numbers of the lower and upper magnon bands (see Figs.~\ref{fig:dispersion_DMI}(a)--(\mw{k})). \pmo{Reversing} \mw{the sign of the \ac{DMI} reverses the role of the two phonon handednesses.}

An experimental signature of this topological gap reversal can be found in transverse thermal transport. The anomalous thermal Hall conductivity $\kappa_{xy}$ relates a temperature gradient to a transverse heat current via $j_x = -\kappa_{xy} \partial_y T$. Within linear response theory, $\kappa_{xy}$ is directly linked to the Berry curvature \cite{Zhang2016,Matsumoto2014,Murakami2017,Klogetvedt2023},
\begin{align}
    \kappa_{xy} = -\frac{k_\mathrm{B}^2 T}{\hbar \mathcal{A}} \sum_{\vec{k}n} c (f_{\vec{k}n}) \Omega_{\vec{k}n}.
\end{align}
Here, $\mathcal{A}$ denotes the system area, $f_{\vec{k}n}$ is the Bose-Einstein distribution function, and $c(x)= (1+x)(\ln \frac{1+x}{x})^2-(\ln x)^2 - 2 \mathrm{Li}_2(-x)$, with $\mathrm{Li}_2(x)$ being the second order polylogarithm function.

Figure~\ref{fig:dispersion_DMI}(\mw{l}) shows the magnon anomalous thermal Hall conductivity for clockwise and counter-clockwise phonon driving. Its behavior closely follows that of the gap $\Delta E$: clockwise phonons enhance $\kappa_{xy}$, while counter-clockwise phonons suppress it and ultimately induce a sign reversal at sufficiently large driving amplitudes.

\mw{We note that the lattice displacements required to completely close and reopen the topological magnon gap in CrI$_3$ are comparatively large, reflecting the already \pmo{sizable} equilibrium gap ($\SI{4}{\milli\electronvolt}$). At such amplitudes, anharmonic effects and structural instabilities may become relevant. The results presented here should therefore be viewed as a proof of principle. We expect that materials with substantially smaller intrinsic magnon gaps, such as CrCl$_3$~\cite{Klein2019}, could exhibit the same phonon-induced topological engineering at considerably smaller lattice displacements.
}

\textit{Conclusion.}
We have demonstrated that magnon band structures can be systematically engineered by driving lattice vibrations, establishing {THz-driven} phonons as an effective and symmetry-selective control knob for magnon topology. Using an \textit{ab initio}-based spin-lattice coupling framework combined with Floquet theory, we showed that circular and elliptical phonons carrying finite \acf{PAM} break time-reversal symmetry for magnons and thereby induce topological gaps in otherwise gapless spectra. In contrast, linearly polarized phonons do not modify the magnon dispersion due to symmetry constraints \mw{within the nonrelativistic spin-lattice coupling considered here}.

A central result of this work is that both the magnitude and topology of magnon bands are directly governed by the \ac{PAM}: the induced gap scales with $|\mathrm{PAM}|$, while the associated Chern numbers are determined by its sign. This enables handedness-selective control, where reversing the sense of lattice rotation switches the Berry curvature and topological invariants. We further showed that only specific phonon symmetries (such as $E_g$ modes in CrI$_3$) modulate intersublattice exchange interactions and can thus modify the magnon spectrum, whereas others (e.g., $E_u$ modes) remain ineffective.

Including \acf{SOC} via \acf{DMI}, we find that phonon driving allows continuous tuning of an already existing magnon gap, including its closure and reopening with inverted topology, depending on \ac{PAM}. This behavior is directly reflected in experimentally accessible observables such as the anomalous thermal Hall conductivity. \mw{We note that relativistic spin-lattice coupling, describing phonon-induced or -modulated \ac{DMI} and anisotropies, provides a subleading correction to this mechanism and enables linearly polarized phonons to modify the gap through an effect that is quadratic in \ac{SOC} \cite{SM}.}

Importantly, the theoretical framework is robust: we explicitly \mw{benchmarked the lowest-order van Vleck expansion against the numerically converged Floquet spectrum obtained in an extended Hilbert space framework~\cite{SM}, finding excellent agreement and confirming the reliability of the effective Hamiltonian}. While demonstrated for CrI$_3$, the underlying mechanism relies only on general symmetry principles and spin-lattice coupling, and is therefore expected to be broadly applicable. Promising candidate systems include other two-dimensional van der Waals magnets (such as CrBr$_3$~\cite{Ghazaryan2018}, CrCl$_3$~\cite{Klein2019}, or Fe$_3$GeTe$_2$~\cite{Fei2018}) as well as magnetic insulators with strong spin-lattice coupling and accessible optical phonon modes. These findings open a route toward dynamical control of magnon topology using tailored phonon excitations.

\acknowledgements
\textit{Acknowledgements.} M.W. and P.M.O. acknowledge funding from the German Research Foundation (Deutsche Forschungsgemeinschaft) through CRC/TRR 227 “Ultrafast Spin Dynamics” (Project MF, Project ID No. 328545488). This work was further supported by the Swedish Research Council (VR) and the Knut and Alice Wallenberg Foundation (Grants No. 2022.0079 and No. 2023.0336). We acknowledge funding from the European Union’s HORIZON EUROPE, under Grant Agreement No. 101129641, “OBELIX”. The calculations were partially supported by resources provided by the National Academic Infrastructure for Supercomputing in Sweden (NAISS) at NSC Linköping, partially funded by VR through Grant No. 2022-06725.


\begin{thebibliography}{57}
\bibitem [{Pau()}]{PauliMatrices}%
  \BibitemOpen
  \href@noop {} {}\bibinfo {note} {{ We define $\sigma_0=
  \left(\begin{smallmatrix} 1 & 0 \\ 0 & 1 \end{smallmatrix}\right)$,
  $\sigma_1= \left(\begin{smallmatrix} 0 & 1 \\ 1 & 0
  \end{smallmatrix}\right)$, $\sigma_2= \left(\begin{smallmatrix} 0 & -i \\ i &
  0 \end{smallmatrix}\right)$, and $\sigma_3= \left(\begin{smallmatrix} 1 & 0
  \\ 0 & -1 \end{smallmatrix}\right)$}}\BibitemShut {NoStop}%
\end{thebibliography}
%

\end{document}